\documentclass[preprint,showpacs,preprintnumbers,amsmath,amssymb,prl]{revtex4}

\usepackage{epsfig}    
\usepackage{dcolumn}
\usepackage{bm}

\def\eva3{$eV$$\cdot$\AA$^3$}               
\def\vs{{\em vs.\/}}                    
\def\*#1*/{}                        

\def\deg{\nobreak\hbox{\hskip 2.0 truept\hbox{$^\circ$}}}

\def\iio{{$\langle 110 \rangle$}}
\def\iii{{$\langle 111 \rangle$}}

\newcommand{\refeq}[1]{Eq.~(\ref{#1})}

\begin{document}

\title{Theory of defects in one-dimensional systems:
       the case of Al in Si nanowires}


\author{Riccardo Rurali}
\affiliation{Departament d'Enginyeria Electr\`onica, 
             Universitat Aut\`onoma de Barcelona, 
             08193 Bellaterra, Barcelona, Spain}

\author{Xavier Cartoix\`a}
 \email{Xavier.Cartoixa@uab.cat}
\affiliation{Departament d'Enginyeria Electr\`onica,
             Universitat Aut\`onoma de Barcelona,
             08193 Bellaterra, Barcelona, Spain}

\date{\today}

\begin{abstract}
The energetic cost of creating a defect within a host material is given by the {\em formation energy}. Here we present a formulation allowing the calculation of formation energies in one-dimensional nanostructures, which overcomes the difficulties involved in defining the chemical potential of the constituent species and the possible passivation of the surface. We also develop a formula for the Madelung correction for general dielectric tensors and computational cell shapes.
We apply this formalism to the formation energies of charged Al impurities in silicon nanowires, obtaining concentrations significantly larger than in their bulk counterparts. 
\end{abstract}

\maketitle




The study of the energetics of the formation of defects is a very well-developed 
topic in bulk semiconductor physics~\cite{PantelidesRMP78,FaheyRMP89,
SeebauerMSERB06,DraboldTDS07}, but it is considerably
less mature in nanostructures. A sensible definition of
the chemical potential, the possible presence of passivating agents
on the surface and a proper treatment of the defect charge
state are some of the issues that prevent the extension of the
standard theory from being straightforward.
The formation energy of a defect~\cite{ZhangPRL91} is a quantity of paramount technological importance; it determines the structural configuration and the charge state that a given set of impurities will favor, and it is used in the computation of impurity equilibrium concentrations~\cite{ZhangPRL91,ZhangPRB93,VandeWallePRB94}, solubilities~\cite{VandeWallePRB93,LuoPRB04}, diffusivities~\cite{FaheyRMP89,StumpfPRL94}, dopant compensation mechanisms~\cite{LaksPRB92}, etc. In addition, the calculation of the formation energy is required whenever a comparison between configurations with different number of atoms/chemical species is wanted.


In bulk host materials, formation energies are calculated according to the well-established theory due to Zhang and Northrup~\cite{ZhangPRL91}, where they are formulated in terms of the chemical potentials of the constituent species and the total energy of the system with the impurities. 
On the other hand, for one-dimensional (1D) semiconductor systems, the nonequivalence of the different constituent atoms in, say, a silicon nanowire (SiNW), in addition to surface passivation issues, render the straightforward application of the Zhang-Northrup formalism troublesome. In particular, the definition of the chemical potential of the atomic species involved is ill-defined. This difficulty is considerably lessened in the study of nanotubes, where each atom is equivalent to the others. This has allowed initial calculations of defect formation energies in C and BN nanotubes~\cite{BaierlePRB01,PiquiniNanotech05}, but analogous calculations for semiconductor nanowires have been missing so far.
Additionally, the most stable configuration of a defect in a semiconductor may have a charge state different from zero, depending on the doping condition of the material. In a periodic boundary condition (PBC) formalism, a finite net charge in the simulation cell would give rise to a divergent Coulomb energy because of the interaction with its periodic images. While the correct procedure for the removal of this contribution to the total energy is still a matter of debate, recent reports indicate that the uniform background charge~\cite{LeslieJPCM85,MakovPRB95} and the local-moment counter charge~\cite{SchultzPRL00} yield similar results~\cite{WrightPRB06} for bulk materials. However, a treatment for charged defects in one-dimensional systems is lacking so far. 

In this Letter we propose a framework for the calculation of formation 
energies of neutral and charged point defects in 1D systems. 
As a case study we discuss the formation of Al point
defects, which can provide {\it p}-doping and can be found as 
contaminant from Al-catalyzed growth process~\cite{WangNatureNano07}, 
thus conveying a considerable technological interest.
Specifically, we will deal with substitutional and interstitial defects 
at different radial positions in \iio\ and \iii\ SiNWs of 1.0 and 1.5~nm 
diameter, identifying whether there is a tendency to surface segregation
and their most stable charge state for different doping conditions.

{\em Formation energy in bulk --} According to the well-established 
theory due to Zhang and
Northrup~\cite{ZhangPRL91}, the formation energy of
a charged defect in a semiconductor $\Delta E^f$ is given by
\begin{equation}
\Delta E^f = E^D_{tot} - E^{pure}_{tot} - \sum n_{i} \mu_{i} + q ( \varepsilon_v + \mu_e ) ,
\label{eq:ZhangNorthrup}
\end{equation}
where $E^D_{tot}$ ($E^{pure}_{tot}$) is the total energy of the defective (clean) system, $n_i$ is positive (negative) for atoms added to (removed from) the clean system, $\mu_i$ is the chemical potential of the reservoir supplying the impurities, $\varepsilon_v$ is the top of the valence band of the clean host and $\mu_e$ is the chemical potential for electrons.

{\em Chemical potential --} Extending the use of this formalism to one-dimensional
nanostructures involves some subtleties related to the definitions
of the chemical potential, especially in the case of vacancies, substitutionals and self-interstitials. When a vacancy is formed in bulk, the removed atom is implicitly assumed to be added to the crystal; thus the use of the bulk chemical potential is justified. For a NW, the displaced atom can be added to multiple nonequivalent positions, as opposed to bulk, where all lattice sites are equivalent. This induces an ambiguity with respect to the choice of the chemical potential, but even if somehow a preferred site for the displaced atom were selected, the calculation of its contribution to the total energy would be ill-defined from a computational point of view. In addition, if the NW is passivated, it is not clear how many passivating atoms or fraction thereof should be assigned to the displaced atom. Clearly, any attempt to use \refeq{eq:ZhangNorthrup} directly will not be able to deal with the possibility that the NWs may be surface passivated.

\begin{figure}
\centering
\epsfig{file=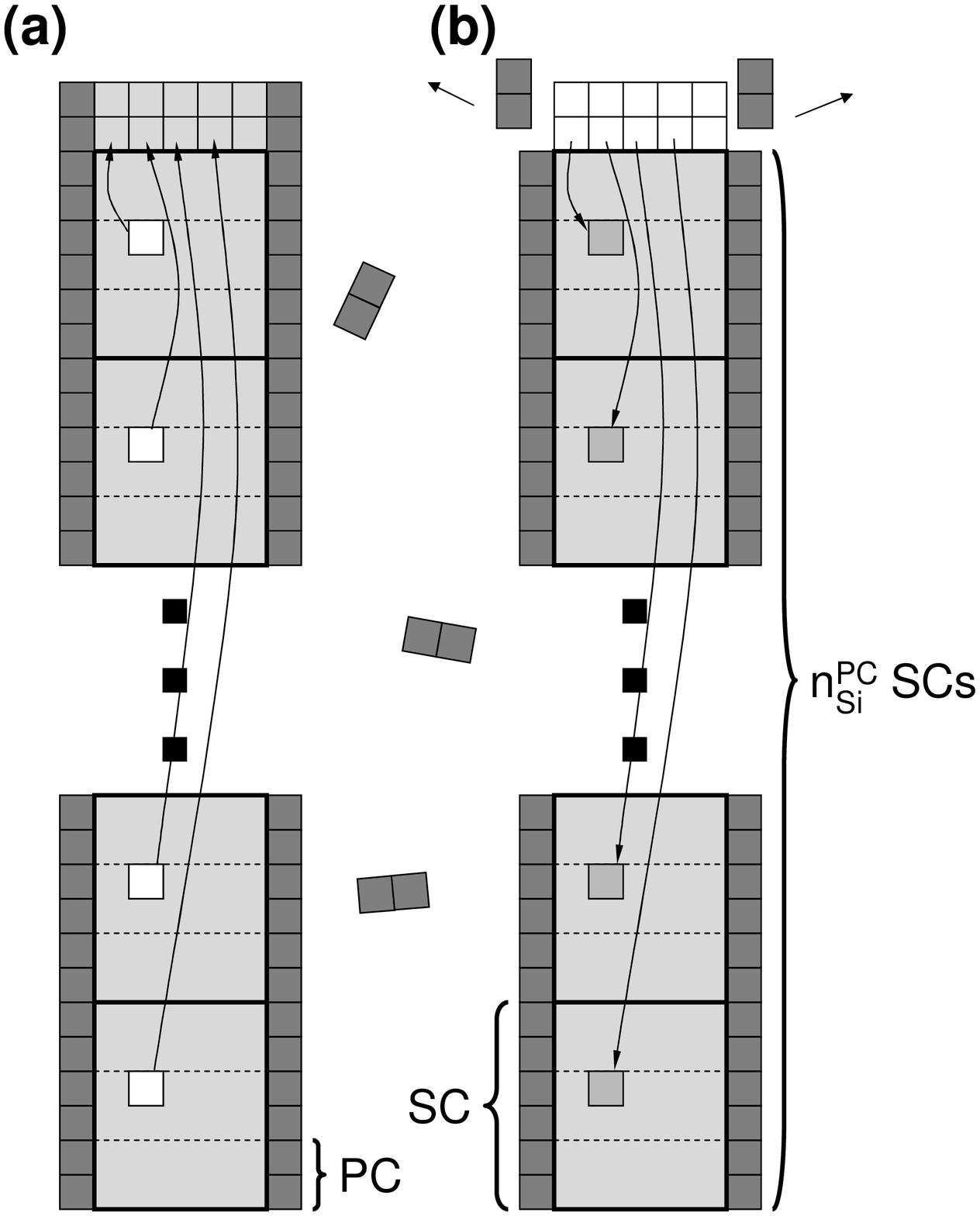, width=0.70\linewidth, clip=, angle=0}
\vskip 0cm
\caption{Construction for dealing with defects in 1D semiconductor structures. (a) Vacancies and substitutionals: we create as many defects as necessary for adding an extra primitive cell to the wire. (b) Self-interstitials: we create as many defects as necessary for removing a primitive cell from the wire.}
\label{fig:hyper}
\end{figure}

We can circumvent these issues by considering the formation of $n^{PC}_{Si}$ defects instead of a single defect (see Fig.~\ref{fig:hyper}), where $n^{PC}_{Si}$ is the number of silicon atoms in a NW primitive cell---for the sake of simplicity we will refer to Si atoms, meaning any host material. These displaced atoms are then added at the end of the NW to form an extra primitive cell with the needed passivating agents. We notice that this is equivalent to defining a chemical potential of the wire primitive cell.

With the above or analogous constructions, it is easy to see that a general expression for the formation energy of a given type of defect will be
\begin{multline}
\Delta E^f = E^D_{tot} - N E_{NW}^{PC} - \sum n_{X} \mu_{X} - \\
  \frac{n_{Si}}{n^{PC}_{Si}} ( E_{NW}^{PC} - n^{PC}_H \mu_{H})  + q ( \varepsilon_v + \mu_e ),
\label{eq:EF_NW}
\end{multline}
where $N$ is the number of primitive cells used in the clean system calculation, $E_{NW}^{PC}$ is the energy of the NW clean primitive cell, satisfying $E^{pure}_{tot} = N E_{NW}^{PC}$, $n_X$ is the number of non-host atoms of species $X$ added to the clean system, $\mu_X$ is their chemical potential, $n_{Si}$ is the number of Si atoms involved in the defect formation (e.g. +1 for a self-interstitial), $n^{PC}_{H}$ is the number of passivating atoms in a NW primitive cell (taken to be hydrogens) and $\mu_H$ is the corresponding chemical potential. All quantities appearing in \refeq{eq:EF_NW} are now well defined and easily extracted from a total energy calculation. Note as well that, in the bulk limit, $n^{PC}_{Si} \rightarrow \infty$ and \refeq{eq:ZhangNorthrup} is recovered.

{\em Charged defects --} The study of charged defects in one-dimensional nanostructures in a PBC formalism must overcome various particularities not present in the bulk case, all arising from the fact that a dielectric {\em tensor} $\bar{\bar{\epsilon}}$ will be needed for the correct description of the interaction between the different instances of the charged defect.

The usual procedure for dealing with these effects in bulk materials consists in using a neutralizing jellium background to
recover the charge neutrality condition, and then correct {\em a posteriori} for the spurious terms arising in the total energy by means of a Madelung correction~\cite{Ziman} divided by the value of the (isotropic) macroscopic dielectric constant of the host material~\cite{MakovPRB95}.

In the case of a NW, the numeric value of the Madelung constant will depend on the relation between the lattice parameters and the chosen $\bar{\bar{\epsilon}}$, and thus it cannot be looked up in tables. Starting from the solution to the Poisson equation within a homogeneous, anisotropic medium~\cite{FischerauerIEEEUFFC97}, and following analogously to the standard procedure~\cite{Ziman}, one can easily obtain the following expression for the Madelung constant in the general case
\begin{multline}
 \alpha = \sum_{ {\bf R}_i } \frac{1}{\sqrt{ \textrm{det } \epsilon} }
   \frac{\textrm{erfc}( \gamma \sqrt{ {\bf R}_i \cdot \epsilon^{-1} \cdot {\bf R}_i } )}{\sqrt{ {\bf R}_i \cdot \epsilon^{-1} \cdot {\bf R}_i }} + \\
   \sum_{ {\bf G}_i }
    \frac{4 \pi}{V_c} \frac{ \exp( {\bf G}_i \cdot \epsilon \cdot {\bf G}_i / 4 \gamma^2 ) }{ {\bf G}_i \cdot \epsilon \cdot {\bf G}_i } -
   \frac{2 \gamma}{ \sqrt{ \pi \textrm{det } \epsilon}  } - \frac{\pi}{V_c \gamma^2} ,
\end{multline}
where the sum over ${\bf R}_i$ (${\bf G}_i$) extends over all vectors of the direct (reciprocal) lattice except for zero, $\gamma$ is a suitably chosen convergence factor and $V_c$ is the volume of the primitive cell.
This approach, besides dealing with generic dielectric tensors, 
allows us also to easily tackle non-conventional cell shapes. This 
is a very common situation in 1D systems, where the 
axial lattice parameter obeys the periodicity of the crystal 
structure, while the transverse dimensions are normally 
much larger. This allows a proper buffer vacuum to avoid spurious 
interactions with the image neighboring system~\cite{MadelungSign}.
Concerning the choice of the dielectric tensor $\bar{\bar{\epsilon}}$, one would be tempted to use, for example, the modified Penn model~\cite{TsuJAP97} in order to obtain an approximation to the value of the dielectric constant for directions perpendicular to the growth axis. This would be indeed the correct approach if we were to study the physics of excitons in the NW. However, it must be kept in mind that in the case at hand two images corresponding to different instances of the NW will interact mainly through vacuum if enough buffer space is left. Thus, we will take the dielectric tensor to be $\textrm{diag}\left( 1, 1, \epsilon_r \right)$, where 
$\epsilon_r$ is the bulk dielectric constant of the constituent material of the NW.

\begin{figure}
\centering
\epsfig{file=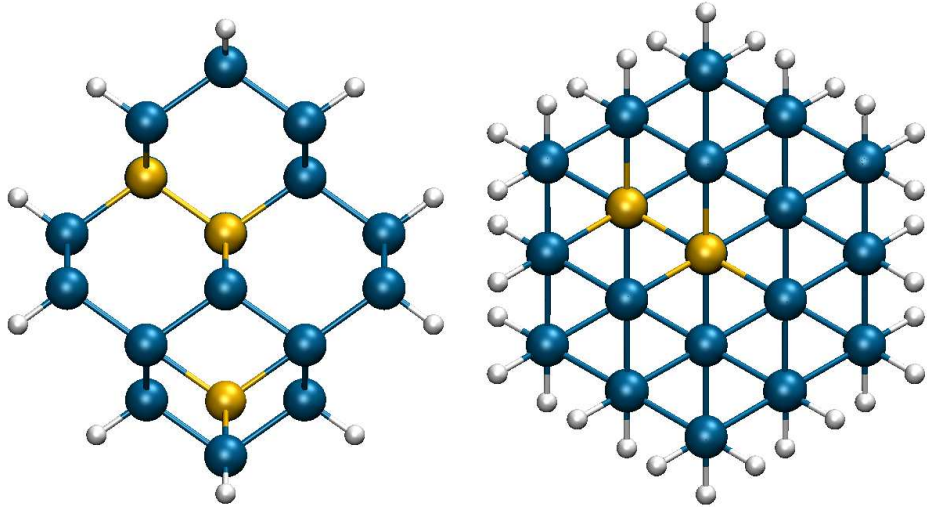, width=0.80\linewidth, clip=, angle=0}
\vskip 0cm
\caption{Cross section view of the 1.0~nm \iio\ (left) and \iii\ 
         SiNWs (right) studied. The substitutional and interstitial 
         Al defects considered are shown in light orange spheres 
         (blue and white spheres representing Si and H atoms, 
         respectively). Notice that that the tetragonal interstitial
         is not visible in the \iii\ SiNWs because it lies along
         the wire axis and it is covered by host atoms and by the 
         innermost substitutional Al defect.}
\label{fig:struct}
\end{figure}

{\em Al defects in \iio\ and \iii\ SiNWs --} 
We apply the formalism described above to the study of
Al point defects in 1~nm and 1.5~nm SiNWs grown along 
the \iio\ and \iii\ axes. Two reasons make Al impurities a very 
interesting case study: (i)~group-III elements can be efficient 
{\em p}-type dopants for silicon, and  the use of Al for doping 
in nanowires has indeed been proposed~\cite{DurgunPRB07}; 
(ii)~Al has proven to be a feasible alternative to Au as a catalyst
for the epitaxial growth of SiNWs~\cite{WangNatureNano07},
having the considerable advantage of not introducing undesired
mid-gap states that can act as traps, 
and requiring lower growth temperatures.
We calculate the total energy within density-functional theory,
as implemented in the {\sc Siesta} package~\cite{SolerJPCM02},
using norm-conserving pseudopotentials, an optimized 
double-$\zeta$ basis set~\cite{AngladaPRB02} and the 
spin-polarized version of the generalized gradient approximation 
(GGA)~\cite{PerdewPRL96} for the exchange-correlation energy. 
The supercell size was chosen to guarantee a separation of 
$\sim$23.7~\AA\ ($\sim$28.3~\AA) between the impurity and its 
periodic image along the \iio\ (\iii) axis, while the transverse
dimensions were held at 50~\AA. The atomic positions were
relaxed until the forces on all the atoms were lower than
0.02~eV/\AA\ and the axial lattice parameter was optimized
for the pristine wire for each growth orientation~\cite{VoPRB06}.
If an Al nanoparticle is used to catalyze the growth reaction,
the nanoparticle will be the main source of Al contaminants
in the SiNW. Therefore, we define the Al chemical potential 
with respect to an Al  particle of approximately the same diameter 
of the wire.

\begin{figure}[t]
\centering
\epsfig{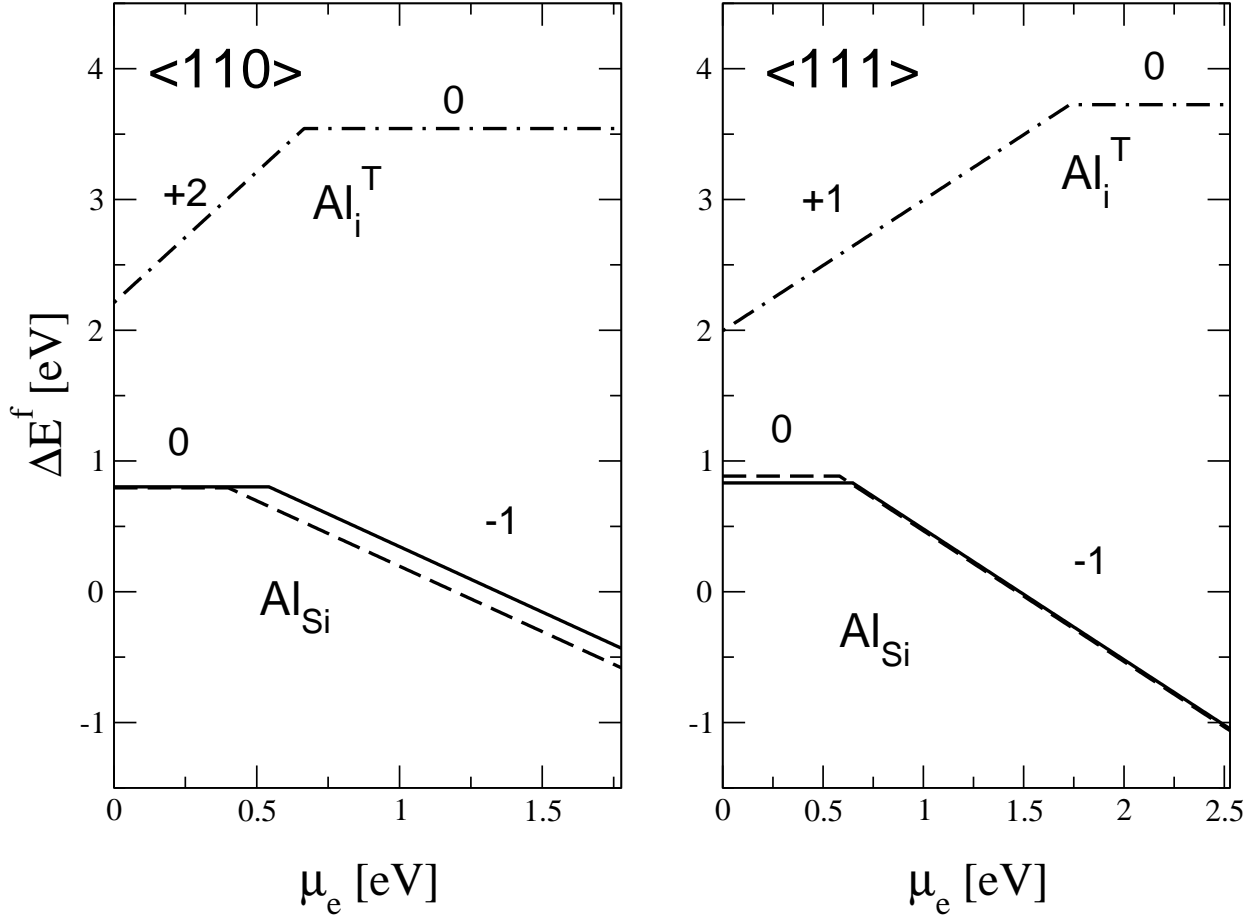}
\vskip 0cm
\caption{Formation energy of Al substitutional (Al$_\text{Si}$) and Al
         tetragonal interstitial (Al$_i^T$) in \iio\ and \iii\ 1.0~nm SiNWs.
         Substitutional defects have been studied at two different radial
         coordinates: at the center of the wire (continuous line) and
         at an equidistant position from the center and the surface
         of the wire (dashed line).}
\label{fig:eform_1.0}
\end{figure}

At first we have considered Al substitutional impurities, Al$_\text{Si}$,
at two different radial coordinates, and the tetragonal
interstitial, Al$_i^T$, in 1~nm \iio\ and \iii\ SiNWs
(see Fig.~\ref{fig:struct}).
As expected, Al is a single acceptor, with substitution at a host
lattice site being largely favored over the interstitial position
(see Fig.~\ref{fig:eform_1.0}).
Despite the thin diameter of the wire, the formation energy of
the Al$_\text{Si}$ is hardly sensitive to the change in the radial
coordinate of the lattice site; the only noticeable change is a slight
lowering of the $0/-$ occupation level of the \iio\ wire when the Al 
substitutes moves outwards (dashed line in Fig.~\ref{fig:eform_1.0}).  
We have not observed any marked tendency to surface segregation,
with the only exception of {\em n}-doped \iio\ wires, where the 
center Al$_\text{Si}$ releases 0.15~eV when moving to more
external position.  It should be noted, however, that Fern\'{a}ndez-Serra 
{\em et al.} showed that the segregation energies rapidly increase in 
presence of surface defects~\cite{Fernandez-SerraPRL06}, a situation 
not considered here.
The most important fact, however, is that the formation energy
of the Al$_\text{Si}$ is negative for {\em n}-type doping
condition, meaning that Al is more stable as a point defect in the wire
rather than in the catalyst nanoparticle. Hence, {\em n}-type doping 
in presence of an Al catalyst might be impractical, as it results
in easy Al incorporation in the wire, leading to uncontrolled 
compensation. A role is certainly played by the reduced value 
of the Al chemical potential $\mu_\text{Al}$ in the nanoparticle 
with respect to bulk $\alpha$-Al (see Eq.~\ref{eq:EF_NW}).
From the behavior of the formation energy \vs\ the Fermi level,
we can infer that choosing properly the substrate doping
allows the control of the wire dopant concentration.
Straightforward application of the Zhang-Northrup formalism,
using bulk derived chemical potentials, results in an
underestimation of the formation energies of $\sim$~0.1~eV.

\begin{figure}[t]
\centering
\epsfig{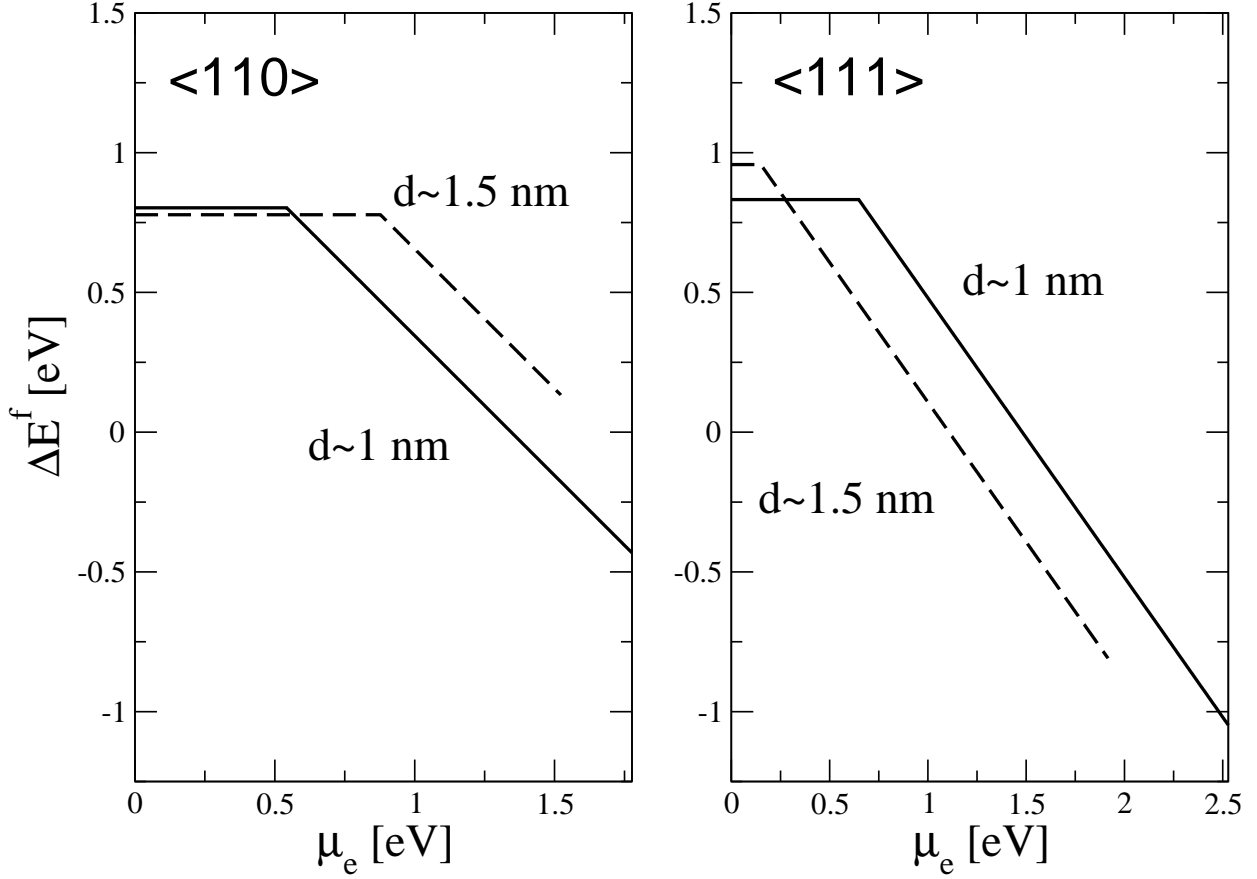}
\vskip 0cm
\caption{Formation energy of Al substitutional (Al$_\text{Si}$) point
         defect in \iio\ and \iii\ SiNWs with a diameter of 1.0 and 1.5~nm.
         The Al$_\text{Si}$ is at the center of the wire (continuous line
         in Fig.~\ref{fig:eform_1.0}). The electron chemical potential
         $\mu_e$ spans the NW band-gap, which is larger for the thinner
         wires.}
\label{fig:eform_1.5}
\end{figure}

In an attempt to address the formation energy \vs\ the  
wire diameter, we have studied the Al$_\text{Si}$ in 1.5~nm
SiNWs, restricting ourselves to substitution at the innermost 
lattice site.  However, we have found that at such small diameters 
the dependence is highly non-trivial, and a clear trend can hardly
be established (see Fig.~\ref{fig:eform_1.5}). 
On the one hand the formation energy of the neutral defect varies
as $\mu_\text{Al}$ of the nanoparticle approaches the bulk value; 
on the other hand, the energy at which the Al$_\text{Si}^-$ becomes 
dominant changes due to the quantum confinement effect.

The impurity concentration $N_i$ is related to the formation energy 
$E_f$ through $N_i = N_s \exp{(-E_f/k_B T)}$, where $N_s$ is the 
concentration of available sites. We estimate, for $T=490$\deg C, an
{\em n}-type doped Si substrate, and $E_f=0.45$~eV as obtained from 
Fig.~\ref{fig:eform_1.5}, that the Al concentration for a 1.5~nm \iii\ 
SiNW is 5.33$\times$10$^{19}$ cm$^{-3}$. 
This is in agreement with the strong Al {\em p}-doping of the SiNWs
reported by Wang {\em et al.}~\cite{WangNatureNano07}.
Unfortunately, they can only provide a rather loose upper bound
on the Al concentration, 10\%, due to the limited sensitivity
of their experimental setup. 
We notice that such a concentration alone cannot explain the
observed tapering of the wires based on a catalyst nanoparticle
consumption mechanism. Hence, either uncatalyzed deposition of Si on
the side of the wires~\cite{WangNatureNano07} or catalyst metal
outdiffusion~\cite{HannonNature06,KodambakaPRL06} must be invoked.

Nevertheless, we expect SiNWs to be able to sustain larger
Al concentrations than their bulk counterparts. Let us consider
the vapor-solid-solid growth of {\em p}-type layers on top
of bulk Si. If the substrate is {\em n}-doped, Al
is easily incorporated in the beginning of the growth process.
However, as the impurity concentration increases,
a space charge zone will develop, bringing the Fermi level
closer to the valence band, 
and the formation energy of Al point defects will become larger.
In these conditions we predict an Al concentration of 2.3$\times$10$^{17}$ 
cm$^{-3}$. In the case of SiNWs this compensation mechanism is not 
expected to be as efficient as in thin film growth, because the
Al impurities incorporated in the reduced NW volume will not
significantly move the Fermi level of the whole system,
allowing larger Al incorporation.

In summary, we have presented a generalized formalism 
that allows to calculate the formation energy of a defect
in one-dimensional semiconductor systems.
We avoid
using bulk derived quantities and we introduce the unambiguously
defined chemical potential of the nanowire primitive cell. The Madelung
correction is extended to the case of arbitrary cell shape and
dielectric tensor. We apply this formalism to the study of acceptors
in silicon nanowires grown catalytically, focusing on the stable
Al substitutional defects. The relatively low formation energy
for {\em p}-doping conditions further decreases
as the Fermi energy moves upwards and it finally becomes negative, 
thus leading to indiscriminate incorporation of Al from the 
catalyst nanoparticle in absence of compensation mechanisms. 
We have calculated the Al concentration for silicon nanowires
grown with the vapor-solid-solid mechanism, predicting an Al
solubility at least one order of magnitude larger than in bulk. 

\begin{acknowledgments}
XC and RR acknowledge financial support from Spain's Ministry of
Education and Science Ram\'{o}n y Cajal program and funding under
Contract No. TEC2006-13731-C02-01.
\end{acknowledgments}

\bibliography{./complete}

\begin{thebibliography}{30}
\bibitem[{Mad()}]{MadelungSign}
\bibinfo{howpublished}{We note that changing the aspect ratio of the simulation
  cell, the Madelung correction can even change sign. Hence, at a certain
  aspect ratio the correction is not needed. This is only an apparent
  advantage, beacause the Madelung correction has to be calculated anyway
  previously to find out which this aspect ratio is.}
\end{thebibliography}
\bibliographystyle {apsrev}

\end{document}